\documentclass[onecolumn]{aastex62}

\newcommand{\chtoh} {CH$_3$OH}
\newcommand{\lo}        {${\rm L}_{\odot}$}                  
\usepackage{natbib}
\usepackage{amsmath}	
\usepackage[normalem]{ulem}  

\shorttitle{\chtoh~toward high-mass star-forming cores}
\shortauthors{Hern\'andez-Hern\'andez et al.}

\begin{document}

\title{APEX Millimeter Observations of Methanol Emission Toward High-Mass Star-Forming Cores}

\author{Vicente Hern\'andez-Hern\'andez}
\altaffiliation{Consejo Quintanarroense de Ciencia y Tecnolog\'\i{a}, Av Insurgentes 210, Chetumal, Q.R., Mexico}
\affiliation{Instituto de Radioastronom\'\i{}a y Astrof\'\i{}sica,
Universidad Nacional Aut\'o\-noma de M\'e\-xico,
Apdo. Postal  3--72 (Xangari), 58090 Morelia, Michoac\'an, M\'exico} 

\author{Stan Kurtz}
\affiliation{Instituto de Radioastronom\'\i{}a y Astrof\'\i{}sica,
Universidad Nacional Aut\'o\-noma de M\'e\-xico,
Apdo. Postal  3--72 (Xangari), 58090 Morelia, Michoac\'an, M\'exico} 
\email{S. Kurtz: s.kurtz@irya.unam.mx}

\author{Sergei Kalenskii}
\affil{Astro Space Center, Lebedev Physical Institute, Russian Academy of Sciences}
\author{Polina Golysheva}
\altaffiliation{Sternberg Astronomical Institute of the Moscow State University}
\affil{Astro Space Center, Lebedev Physical Institute, Russian Academy of Sciences}

\author{Guido Garay} 
\affil{Departamento de Astronom\'\i{}a, Universidad de Chile,  
Camino del Observatorio 1515, Las Condes, Santiago, Chile}

\author{Luis Zapata}
 \affil{Instituto de Radioastronom\'\i{}a y Astrof\'\i{}sica,
Universidad Nacional Aut\'o\-noma de M\'e\-xico,
Apdo. Postal  3--72 (Xangari), 58090 Morelia, Michoac\'an, M\'exico}

\author{Per Bergman}
\affil{Dept of Space, Earth and Environment, Chalmers Univ. of Technology,
Onsala Space Observatory, SE-43992 Onsala, Sweden}

\begin{abstract}
We present 247-GHz molecular line observations of methanol (CH$_3$OH) toward
sixteen massive star-forming regions, using the APEX
telescope with an angular resolution of 25$''$. 
The sample covers a range of evolutionary states, including warm molecular cores,
hot molecular cores, and ultracompact HII regions.
The hot cores, all of which include UC HII regions, show rich molecular
line spectra, although the strength of different species and
transitions varies from source to source. In contrast, the warm 
cores do not show significant molecular line emission. Multiple
methanol transitions are detected toward nine of the hot
cores; eight of these had enough transitions to use the rotation
diagram method to estimate rotational temperatures and column
densities. The temperatures lie in the range 104--168 K and column
densities from $3\times10^{16}$ to $7\times10^{18}$\,cm$^{-2}$. Using the
average methanol line parameters, we estimate virial masses,
which fall in the range from 145 to
720 M$_{\odot}$~and proved to be significantly higher than the measured gas masses.
 We discuss possible scenarios to explain the chemical
differences between hot cores and warm molecular cores. One of the observed methanol 
lines, $4_2-5_1$A$^+$ at 247.228 GHz, is predicted to show class II maser emission, 
similar in intensity to previously reported $J_0-J_{-1}$E masers at 157~GHz. We did 
not find any clear evidence for maser emission among the observed sources;
however, a weak maser in this line may exist in G345.01+1.79.
\end{abstract}




\section{Introduction}
\label{sec:intro}
Massive stars ($M>8$\,M\;$_\odot$) play a fundamental role in the evolution of galaxies. 
The large amounts of radiation and mechanical energy they deposit into the interstellar
medium can profoundly affect the star formation rate, either by initiating new star formation
by compressing the interstellar gas or terminating star formation, by heating and dispersing
molecular clouds.
\citep[see, e.g., ][]{2007ARA&A..45..481Zinnecker}.  

Studies show that star-forming regions (SFRs) within Infrared Dark Clouds
\citep[IRDCs, e.g.,][]{2009ApJS..181..360C} contain massive cores with
typical masses of $\sim120$\,M$_\odot$, sizes of $<$0.5\,pc, densities of
$\sim10^{5}-10^{6}$\,cm$^{-3}$, and temperatures of 15--30\,K
\citep[e.g.,][]{2006ApJ...641..389Rathborne}. While most of these
massive cores show no evidence for active star formation (they are often
called cold or quiescent cores;
\citep[e.g.,][]{2006ApJ...641..389Rathborne}, others show bright
24\,$\mu$m emission, high-excitation molecular line emission, and
maser emission. These warm molecular cores (WMC) are thought to be
actively forming massive stars
\citep[e.g.,][]{2004ApJ...610..313Garay,2009ApJS..181..360C}.

Hot molecular cores (HMC) are more-evolved objects than WMC,
with higher temperatures ($>$100\,K), densities
($\sim10^7$\,cm$^{-3}$), and luminosities ($>10^4$\,L$_{\odot}$). They are
compact ($<$0.1\,pc) and have a very rich chemistry. Their molecular
gas is probably heated by embedded massive (proto)stars
\citep[][]{2000prpl.conf..299Kurtz,2005IAUS..227...59C}. HMC can also
host energetic outflows while simultaneously undergoing infall in
their outer regions \citep[e.g.,][]{2009ApJ...694...29O,2014MNRAS.437.3766M}.  
The mass accretion rates in some cases may be sufficient
to quench the development of H{\small II}~regions
\citep[][]{1995Ap&SS.224..173Walmsley,2008A&A...479L..25Z,2009ApJ...694...29O}.  Thus, it
is thought that HMCs precede the development of ultracompact (UC) H{\small II}~regions.

Some molecular species in HMCs show maser emission,
including H$_2$O, H$_2$CO, NH$_3$, OH, and CH$_3$OH.  In particular, CH$_3$OH~
presents a complex spectrum with numerous thermal and/or maser lines. These
lines are sensitive to the molecular gas conditions, allowing a
detailed analysis of physical parameters.  Methanol masers are also
important tracers of dynamical phenomena associated with massive
star formation: they are often associated with molecular outflows,
disks, and shocks \citep[][and references
therein]{2005A&A...429..945M,2006ApJ...638..241E}.

Methanol masers are grouped into two classes \citep[I and
  II;][]{1991ApJ...380L..75M}.  Class I masers are typically  offset from the protostellar position with the more distant masers
  located $\gtrsim$0.1 pc away from them~\citep[][]{2004ApJS..155..149Kurtz}; 
 many of these masers are found to be substantially closer,
  however,~\citep[][]{2018MNRAS.477..507McCarthy}.  Class I masers
are thought to be collisionally pumped
\citep[][]{1992MNRAS.259..203C,2016A&A...592A..31L}. Class II masers
are usually found very near to the protostellar position and are
thought to be radiatively pumped
\citep[][]{2005MNRAS.360..533C}. Methanol maser surveys have shown
that class II masers are exclusively associated with massive star
formation regions (MSFRs)
\citep[][]{2003A&A...403.1095M,2008A&A...485..729X}.  The strongest
class II masers are observed in the centimeter-wave regime, while
weaker lines are detected at 3, 2, and 1~mm
\citep[][]{1995A&A...294..825V,1995ApJ...442..668S,
  2017A&A...606L...6Z}.

\citet[][]{1997MNRAS.288L..39S} and \citet[][]{2005MNRAS.360..533C} developed models to
predict class II masing of methanol transitions in the 6 to 700\,GHz
range. Their models demonstrate that maser
intensity diminishes at higher frequencies due to lesser degrees of
population inversion. To constrain these models it is
useful to determine which high-frequency transitions show maser emission and
the relative intensities of these transitions.
In the millimeter regime, one of the better maser candidates is the
$4_2-5_1$ A$^+$ transition at 247.228\,GHz. 
 Model 3 of \citet[][]{2005MNRAS.360..533C} predicts a
brightness temperature of $10^{6.2}$\,K for this line, approximately
the same as the brightness temperature of known class II masers at 
157\,GHz \citep[][]{1995ApJ...442..668S} and about a factor of 5 lower 
than the brightness temperature of another known Class II maser line, 
$3_0-4_{-1}$~A$^+$ at 107~GHz~\citep{1995A&A...294..825V}. This
model predicts a brightness temperature ratio for the 6.7 and
247.2\,GHz masers of $10^{4.3}$, and hence the predicted ratio of 6.7/247.2\,GHz
flux densities is about 15. Because many 6.7\,GHz masers show flux
densities of hundreds of Janskys, one can expect that 
masers at 247.2\,GHz will have intensities high enough to be easily
detected.

\begin{deluxetable}{lcccccccc}
\tabletypesize{\scriptsize}
\tablewidth{0pt}
\tablecaption{Observed Sources \label{sources} }
\tablehead{\colhead{Source}&\colhead{Short} &\colhead{$\alpha\;$(J2000)}&\colhead{$\delta\;$(J2000)}
     &\colhead{V$_{LSR}$}&\colhead{Distance}&\colhead{$L_{\rm bol}$}&\colhead{}&\colhead{}\\ 
\colhead{}  & \colhead{Name}&\colhead{$^h\;^m\;^s$}&\colhead{$^{o}\;{'}\;{''}$}&\colhead{(km~s$^{-1}$)}
&\colhead{(kpc)}    &\colhead{($10^5$ \lo)} &\colhead{Comments$^{a}$}& \colhead{Refs.} } 
\startdata
G345.01+1.79   &  G345.01   &$16^h56^m47.6^s$&$-$40 14 26  &$-$14        &  2.4    & 0.21    & HMC+UC\,H{\small II} & 12 \\
G351.417+0.645 & N6334~F    &  17 20 53.4   &$-$35 47 01  &$-$10.5       & 1.3$^b$ &  1.5      & HMC+UC\,H{\small II}     & 11,14         \\
G351.446+0.659 & N6334~I(N) &  17 20 55.0   &$-$35 45 05   &$-$3        & 1.3$^b$ & 0.01      & HMC+UC\,H{\small II}      & 1,14        \\ 
G351.776-0.537 & I17233     &  17 26 43.0   &$-$36 09 15   &$-$3        & 1.2     & 0.14      & HMC+UC\,H{\small II}      & 2,3,13   \\ 
G9.621+0.196   & G9.62      &  18 06 14.7   &$-$20 31 32   &$-$3         & 5.2$^b$ &            &HMC+UC\,H{\small II}     &  16        \\
G10.47+0.03    & G10.47     &  18 08 38.2   &$-$19 51 49   & +68         & 8.5$^b$ & 7.0       & HMC+UC\,H{\small II}     & 4,5        \\ 
G10.62--0.38   & G10.62     &  18 10 28.7   &$-$19 55 49   &$-$3        & 5.0$^b$ & 9.2       & HMC+UC\,H{\small II}    & 5,6        \\ 
G20.08-0.14N   & G20.08N    &  18 28 10.3   &$-$11 28 48   & +42         & 12.3    & 6.6       & HMC+UC\,H{\small II}    & 7        \\ 
G24.33+0.11MM1 & G24.33     &  18 35 08.1   &$-$07 35 04   & +53         & 3.8     & 0.4       & WMC              & 10        \\ 
G28.53--0.25MM2& G28.53     &  18 44 15.7   &$-$03 59 41   & +87         & 5.7     & 0.005     & WMC              & 10        \\ 
G29.96--0.02   & G29.96     &  18 46 03.9   &$-$02 39 21   & +97         & 5.2$^b$ & 4.3       & HMC+UC\,H{\small II}   & 8,5        \\ 
G30.97--0.14MM1& G30.97     &  18 48 22.0   &$-$01 48 27   & +79         & 5.1     & 0.04      & WMC              & 10        \\ 
G31.97+0.07MM1 & G31.97     &  18 49 36.6   &$-$00 45 45   & +97         & 5.1     & 0.36      & WMC              & 10,5      \\   
G34.26+0.15    & G34.26     &  18 53 18.5   & +01 14 57    & +58         & 3.7      & 5.2      & HMC+UC\,H{\small II}      & 9        \\ 
G35.204-1.738  & W48        &  19 01 46.2   & +01 13 42    & +45.7       & 3.27$^b$ & 0.08     & UC\,H{\small II}            & 15        \\ 
G45.07+0.13    & G45.07     &  19 13 22.0   & +10 50 53    & +60         & 8.0$^b$  & 11.0     & HMC+UC\,H{\small II}     & 4,5      \\ 
\enddata
\tablecomments{Units of right ascension are hours, minutes, and seconds, and for declination are degrees, arcminutes, 
and arcseconds. Positions, V$_{\rm lsr}$, and luminosities are taken from the cited references. (a) Hot molecular cores 
(HMC), all of them showing ultracompact HII regions (UC\,H{\small II}), and warm massive cores (WMC) associated with infrared dark clouds.
(b) trigonometric parallax distance, taken from~\citep[][and references therein]{2014ApJ...783..130R}. Otherwise, the distance is kinematic and taken from the cited references.\\
 References--- (1) \citet{2014ApJ...784..114C}; (2) \citet{2011AA...530A..12L}; 
(3) \citet{2004AA...426...97F}; 
(4) \citet{2008AA...486..191P}; (5) \citet{2014ApJ...783..130R}; (6) \citet{2005ApJ...630..987S}; (7) \citet{2009ApJ...706.1036G}; 
(8) \citet{2007AA...468..1045B}; (9) \citet{2007ApJ...659..447M}; (10) 
\citet{2006ApJ...641..389Rathborne}; (11) \citet{2018ApJ...854..170H}; (12) 
\citet{2013ApJS..208...11L}; (13) \citet{2008AJ....136.1455Z}; (14) \citet{2016ApJ...832..187B}
(15) \citet{2014MNRAS.440..427R} (16) \citet{2017ApJ...849...25L} 
}
\end{deluxetable}

An additional advantage of selecting the 247.2~GHz line for the observations
is the location of its energy levels. The upper level of the $4_2-5_1$~A$^+$ transition belongs 
to the $K=2\;$A$^+$ ladder while the lower level belongs to the $K=1\;$~A$^+$ ladder. 
Little is known about masers in $K=2 \rightarrow K=1$ transitions of methanol A. Only three regions, 
W3(OH), NGC6334F, and NGC7538, have been found to show maser emission in the $9_2-10_1$~A$^+$ 
transition at 23~GHz~\citep[][and references therein]{2004MNRAS.351.1327C}. A fourth 23~GHz methanol
maser has recently been reported in G358.93$-$0.03, but it has not yet been confirmed by interferometric
observations (O. Bayandina, personal communication).
 \citet{1995ApJ...442..668S} detected
the 156.1~GHz $6_2-7_1$~A$^+$ transition in only one source, W3(OH), and reported
a weak, broad line, which may contain a narrow maser component.  Although these masers
are fairly rare, information about their intensities will help to constrain maser models.

To test the predictions of the Sobolev/Cragg model, we observed 16
MSFRs associated with WMCs, UC\,H{\small II}, or HMC+UC\,H{\small II}-regions, using
the APEX antenna.  The latter sources are HMC that are closely
associated with UC\,H{\small II}~regions.  Fifteen of the sources (all 
but G28.53) show methanol maser emission at 6.7\,GHz, with peak fluxes ranging from 2 to $\sim 5000$\,Jy, and 
a median value of 36\,Jy. In addition, G345.01+1.79 and W48 are the only
two masers at 157 GHz detected by \citet{1995ApJ...442..668S} in the southern
hemisphere; these masers also emit at 107~GHz~\citep{1999MNRAS.310.1077V}; G9.62+0.19 
is the strongest maser at 6.7~GHz, according to~\citet{1991ApJ...380L..75M}
and shows maser emission at 107~GHz \citep{1999MNRAS.310.1077V}, and NGC6334F 
presents strong maser emission at 23~GHz~\citep{2004MNRAS.351.1327C} 
and 107~GHz~\citep{1999MNRAS.310.1077V}. Thus, our target list includes the most 
probable candidates for detecting masers in the $4_2-5_1$~A$^+$ transition, achievable for APEX. 

Our APEX observations and the reduction
procedures are described in Sect.~\ref{sec:observ}.  Results and
discussion are presented in Sect.~\ref{sec:results}, and we make some
concluding remarks in Sect.~\ref{sec:conclus}.

\section{Observations and data reduction}
\label{sec:observ}

In Table~\ref{sources} we list the observed sources. 
The sample is comprised of four WMC, one UC\,H{\small II}~region
and eleven HMC + UC\,H{\small II}-regions. The WMCs were selected from a list of cores 
associated with IRDCs~\citep{2006ApJ...641..389Rathborne}. These four sources have masses
(determined from 1.2\,mm dust continuum emission) above 400\,M$_\odot$, present 4.5 $\mu$m emission
(often associated with shocks and outflows), strong 8.0 $\mu$m
emission, class II CH$_3$OH~maser emission at 6.7~GHz and class~I maser emission
at centimeter wavelengths. The HMC and UC\,H{\small II}-regions were selected from the literature 
and are well-known MSFRs with luminosities greater 
than $10^4$\,L$_\odot$ and all show class II methanol maser emission.
 
The observations were made using the APEX\footnote{Based on
    observations with the Atacama Pathfinder EXperiment (APEX) telescope.
    APEX is a collaboration between the Max Planck Institute for Radio
    Astronomy, the European Southern Observatory, and the Onsala Space
    Observatory. Swedish observations on APEX are supported through Swedish
    Research Council grant No 2017-00648.} 
12-m telescope at Llano de Chajnantor in the Atacama desert of Chile 
(Chilean time project C-086.F-0667B-2010 and Swedish time project O-0100.F-9307A-2017).
The first observing run was performed during 2010 August--December.
We used the SHFI APEX-1 receiver~\citep{2008A&A...490.1157V} and the Fast 
Fourier Transform Spectrometer (FFTS) configured
with two units, each with a bandwidth of 1 GHz and 8192 channels.  We
overlapped the units to achieve 1.8\,GHz total bandwidth, centered at
247228.73\,MHz (1.21\,mm wavelength) which is the rest frequency of
the CH$_3$OH~(4$_{2}-5_{1}$~A$^+$) line. The channel width was 0.15\,km~s$^{-1}$, and 
the total velocity coverage was 2200\,km~s$^{-1}$.  At the observing
frequency of 247 GHz the APEX forward efficiency is 0.95, the main beam
efficiency is $\eta_{\rm mb}=0.75$, and the primary beamwidth is
$\theta_{\rm FWHM} = 25$\rlap.{$''$}2.  
We made single-pointing, beam-switched observations of each source 
for 10 minutes duration (on-source), achieving a typical rms of 45\,mK. We used
RAFGL~2135 and NGC~6302 as pointing sources at different epochs.

\begin{deluxetable}{llcc}
\tabletypesize{\scriptsize}
\tablewidth{0pt}
\tablecaption{Parameters of some detected molecular lines \label{lines17233} }
\tablehead{\colhead{Species}&\colhead{Transition}&\colhead{Frequency}&\colhead{$E_{u}$}\\
\colhead{}                  &\colhead{}          &\colhead{(GHz)}    &\colhead{(K)}
}
\startdata
HC$_3$N     & $27-26~1v7~l=1f$     & 246.5607  & 486      \\  
HC$_3$N     & $27-26~2v7~l=0$      & 247.1146  & 807      \\  
HC$_3$N     & $27-26~2v7~l=2f$     & 247.2099  & 811      \\  
CH$_3$OH    & $21_3A^{-}-21_2$A$^+$ & 245.2235  & 586      \\
CH$_3$OH    & $20_3A^{-}-20_2$A$^+$ & 246.0749  & 537      \\
CH$_3$OH    &  $19_3A^- -19_2$A$^+$ & 246.8735  & 490      \\
CH$_3$OH    &  $16_2-15_3$E         & 247.1619  & 338      \\
CH$_3$OH    &  $4_2-5_1$A$^+$       & 247.2287  & 70       \\
CH$_3$OH    & $18_3A^- -18_2$A$^+$  & 247.6110  & 446      \\
CH$_3$OH    &  $12_{-2}-13_{-3}$E~ 
                           $v_t=1$  & 247.8402  & 545      \\  
CH$_3$OH    &  $23_1-23_0$E         & 247.9681  & 661      \\ 
CH$_3$OH    &  $17_3A^{-}-17_2$A$^+$ & 248.2825  & 405   \\ 
CH$_3$OH    &  $16_3A^{-}-16_2$A$^+$ & 248.8855  & 365   \\ 
SO$_{2}$    & $15_{2}-15_{1}$      & 248.0574        & 119       \\  
SO$_{2}$    & $31_{9}-32_{8}$      & 247.1697        & 654       \\  
$^{34}$SO   & $5_3-5_2$            & 247.4403        & 35          \\  
$^{34}$SO   & $4_1-4_2$            & 246.6861        & 30          \\  
CH$_3$CH$_2$CN &  $28_{1}-27_{1}$  & 248.0425        & 176        \\  
CH$_3$CH$_2$CN &  $27_{3}-26_{3}$  & 246.5487        & 174        \\  
CH$_3$CH$_2$CN &  $28_{2}-27_{2}$  & 246.4219        & 177        \\  
\enddata
\end{deluxetable}

Several of the more promising sources for maser detections at 247~GHz
were not scheduled during the first run. Therefore, we performed
additional observations in August 2017. We used the APEX-1 receiver and the Extended 
Fast Fourier Transform Spectrometer (XFFTS) configured with both spectral units,
overlapped by 1000 MHz, giving  a usable bandwidth of $\sim$ 4 GHz, and allowing 
the search for a number of additional molecular transitions.
The observational technique and the time per source were the same as in the first
observing run. 

The data reduction was done with the CLASS software package\footnote{CLASS is 
part of the GILDAS software package developed by IRAM.}. First, we inspected 
the raw data to eliminate any bad spectra in either of the FFTS units.  The spectra 
were then combined into a single spectral window. The width of the window
was  1.8\,GHz for the 2010 data and 4 GHz for the 2017 data. To fit a baseline,
we selected windows avoiding any obvious line emission and fitted a
polynomial of order 3 or less. The data were smoothed by 6 channels 
(i.e., averaged without overlap) to obtain a velocity resolution of 0.9\,km~s$^{-1}${ }
(0.74\,km~s$^{-1}${} for the 2017 data) and an rms of 0.016--0.018~K
in the final spectra, which we used for further analysis. 
We converted the antenna temperature, $T_{\rm A}^*$, to main-beam
brightness temperature, $T_{\rm mb}$, through $T_{\rm mb} = T_{\rm A}^* / \eta_{\rm mb}$.

\begin{figure}
\centering
\includegraphics[width=0.8\linewidth]{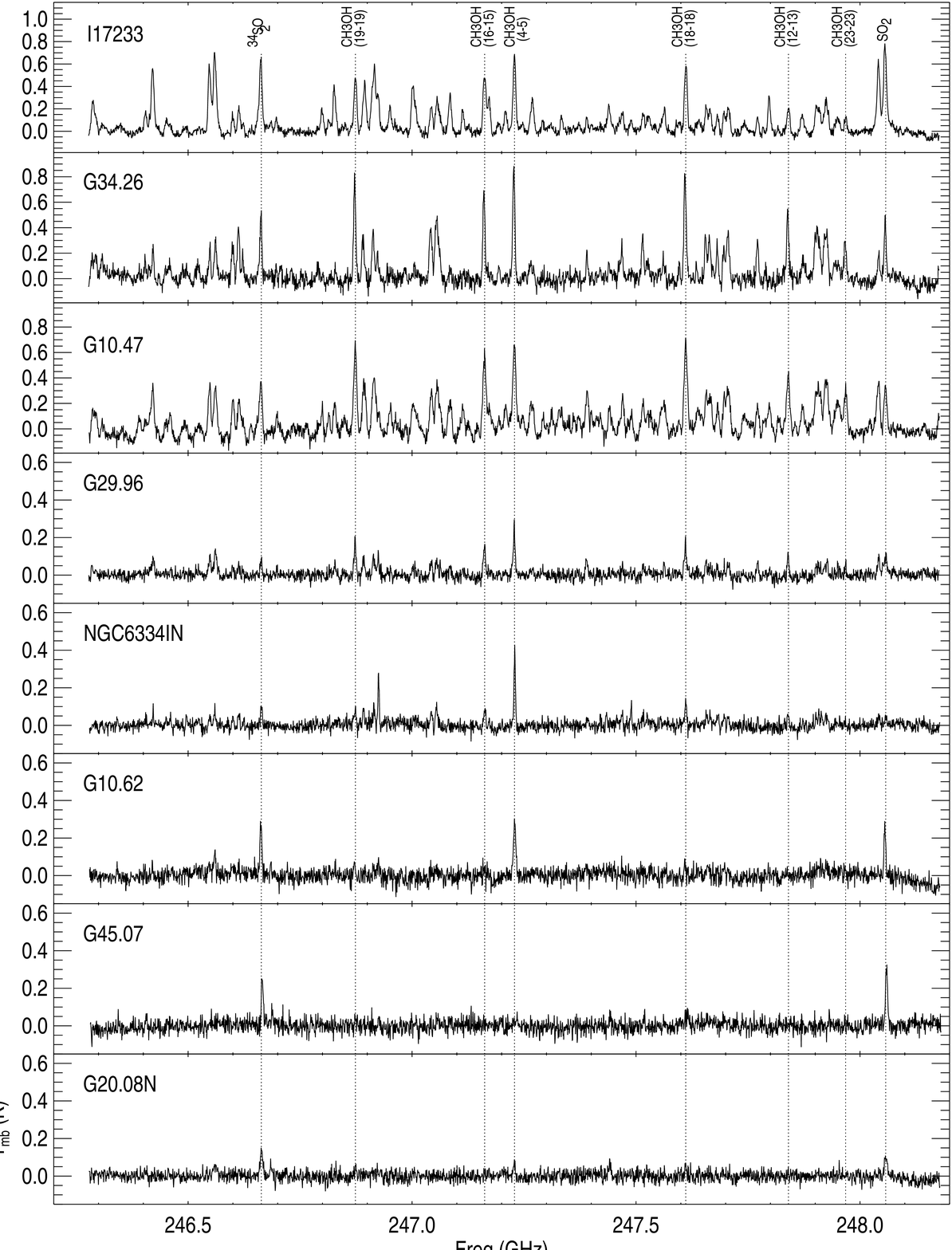}
\bigskip
\caption{Wide-band continuum-free spectra taken with APEX in 2010.}
\label{sp2010}
\end{figure}

\addtocounter{figure}{-1}
\begin{figure}\centering
\includegraphics[width=0.8\linewidth]{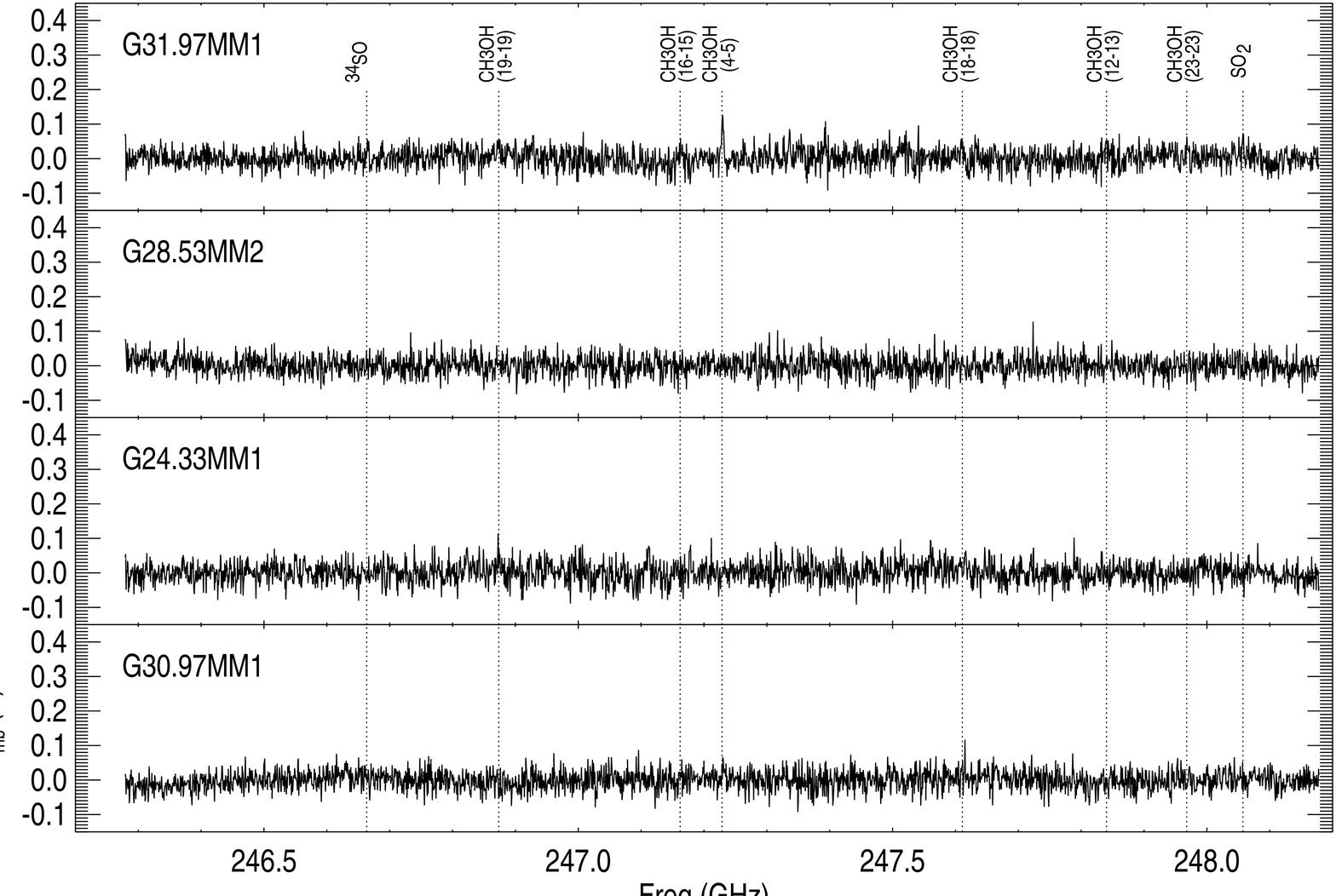}
\medskip
\caption{(continued)}
\end{figure}

\begin{figure*}
\centering
\includegraphics[width=1.0\linewidth]{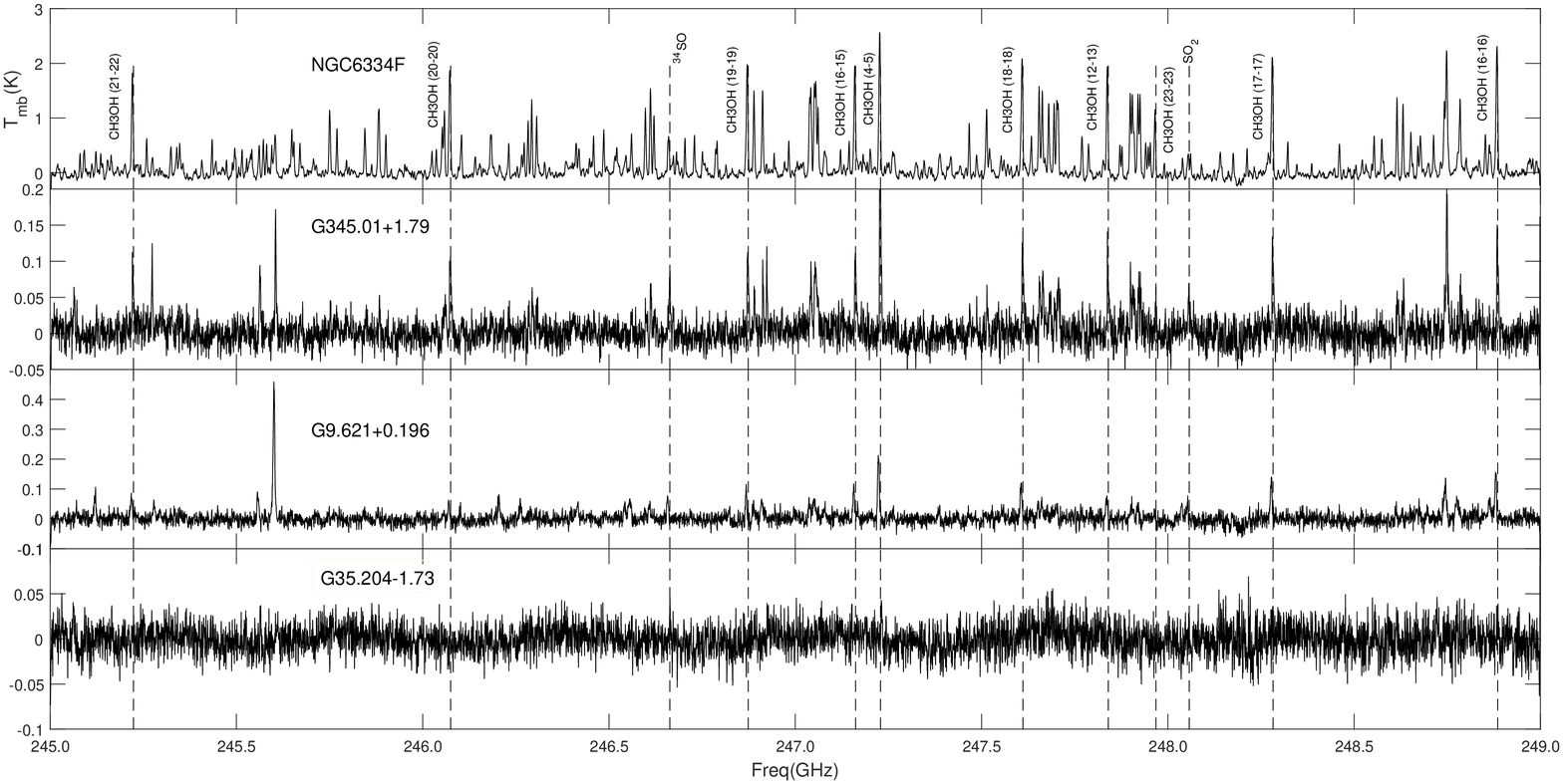}
  \caption{Wide-band continuum-free spectra taken with APEX in 2017.  
We label several CH$_3$OH~lines, the SO$_{2}$ line at 248.05 GHz, and the $^{34}$SO line. 
\label{sp2017}}
\end{figure*}

\section{RESULTS and DISCUSSION}
\label{sec:results}

\subsection{Detection of molecular line emission}

In Figures~\ref{sp2010} and~\ref{sp2017} we show the spectra of 
the sixteen sources.
In Table 2 we list some physical characteristics of the major lines we detected. We used the
SPLATALOGUE\footnote{http://www.splatalogue.net/} website to identify
the main lines in the spectra. We note that the large number of emission 
lines produces blending of some lines in sources such as N6334F, I17233, 
G34.26, and G10.47 (see Figures~\ref{sp2010} and~\ref{sp2017}), which complicates 
the line identification.  High angular resolution ALMA observations would 
be more adequate for this purpose, and also to trace the gas dynamics within the cores.

Four of the 12 sources that harbor HMCs and/or UC\,H{\small II},
NGC6334F, I17233, G10.47, and G34.26,  contain numerous
emission lines from oxygenated, nitrogenated, and sulfurated species,
along with their isotopomers. Distinct molecular species
detected include HC$_3$N, CH$_3$OH, SO$_{2}$, $^{33}$SO$_2$, $^{34}$SO,
CH$_3$OCHO, CH$_2$CHCN and CH$_3$CH$_2$CN. 
The G29.96 and G345.01 HMC+UC\,H{\small II} regions show less-rich spectra, but 
nevertheless present emission in a number of high-energy lines. 
The spectra of NGC~6334~I(N) and G9.62 are even more line--poor than those of
G29.96 and G345.01. 
G10.62, G20.08M, G35.204, and G45.07 show only a few lines in their spectra. 
We note that all the above are HMCs or HMC + UC\,H{\small II}-regions. The four warm
cores, G31.97MM1, G28.53MM2, G24.33MM1, and G30.97MM1, show very little molecular 
line emission.

\subsection{Comparison between sources}

I17233, NGC6334F, G34.26, and G10.47 show notably richer spectra than
other HMCs such as G345.01, NGC~6334~I(N), G9.621, G10.62, G20.08N, G29.96,
and G45.07. This is consistent with the results of
Hern\'andez-Hern\'andez et al. (2014), who studied a sample of 17 hot
cores in methyl cyanide lines and found that the CH$_3$CN line
intensities, and hence the column densities, in G10.62 and G45.07 are
much lower than those in I17233 and G10.47.

In addition, we see
differences in the emission of some molecular lines within these
regions. For example, NGC~6334~I(N) and G10.62 show strong emission
from the CH$_3$OH~$4_2-5_1A^+$ line compared to G45.07 and G20.08N.  On the
other hand, G45.07 and G10.62 show strong emission in SO$_{2}$ (at
248.05\,GHz) and $^{34}$SO (at 246.66\,GHz), while G29.96,
NGC~6334~I(N), and G20.08N are very weak in these lines.  In any case,
for all these sources, we detect the $4_2-5_1A^+$~CH$_3$OH and~$15_1-15_2$ SO$_2$
transitions, which have excitation temperatures of 70 and 120\,K,
respectively.  Differences in the number of transitions and which
molecules are in emission have been reported toward other \mbox{MSFRs}, such
as the Orion hot core and W3(OH)--TW
\citep[][]{1999ApJ...514L..43W,2015ApJ...803...39Q}. Significant
differences in the number and intensity of molecular lines in different SFRs
was found by~\citet[][]{1998A&AS..133...29H}, who divided
the observed sources into line-rich and line-poor. An extensive systematic study
of more than 100 SFRs by~\citet{2017A&A...603A..33G} also showed a similar picture. The
differences have variously been explained as the result of different physical conditions,
different composition of ice-mantles on dust grains, or different
evolutionary stages
\citep[][]{2015A&A...575A..68C,2009AJ....137..406B}.

\bigskip
\startlongtable
\begin{deluxetable}{lcccc}
\tabletypesize{\scriptsize}
\tablewidth{0pt}
\tablecaption{Observed Parameters of CH$_3$OH~lines \label{paramobs}}
\tablehead{\colhead{Source}&\colhead{$T_{\rm mb}$}&\colhead{FWHM}&\colhead{$\int T_{\rm mb}\Delta V$}
&\colhead{V$_{LSR}$}      \\    
\colhead{Transition} &\colhead{(K)} &\colhead{(km~s$^{-1}$)} &\colhead{(K km~s$^{-1}$)}
&\colhead{(km~s$^{-1}$)}  
}
\startdata
G345.01+1.79  &       &                                             &        &  \\
(21--21)  & 0.08      & $5.34\pm 0.60$  & $0.45\pm 0.04$            & -12.30\\
(20--20)  & 0.11      & $6.70\pm 0.50$  & $0.77\pm 0.05$            & -13.56\\
(19--19)  & 0.12      & $5.20\pm 0.36$  & $0.64\pm0.04 $            & -13.31\\
(16--15)  & 0.11      & $5.02\pm0.48$   & $0.58\pm0.05$             & -13.34\\
(4--5)    & 0.32      & $4.11\pm0.18$   & $1.41\pm0.05 $            & -13.00\\
          & 0.08      & $2.48\pm0.59$   & $0.21\pm0.05 $            & -17.62\\
(18--18)  & 0.12      & $6.01\pm0.53$   & $0.79\pm0.06 $            & -13.53\\
(12--13)  & 0.08      & $7.49\pm1.84$   & $0.65\pm0.06 $            & -13.21\\
(23--23)  & 0.05      & $4.77\pm1.08$   & $0.25\pm0.05$             & -13.76\\
(17--17)  & 0.11      & $5.61\pm0.49$   & $0.66\pm0.05$             & -13.32\\
(16--16)  & 0.13      & $5.49\pm0.37$   & $0.74\pm0.05$             & -13.20\\
\hline
NGC6334F  & & & &  \\
(21-21)   & 1.81      & $5.60\pm0.09$   & $10.75\pm0.15$            & -7.06\\
(20-20)   & 2.00      & $6.38\pm0.05$   & $13.57\pm0.09$            & -7.78\\
(19-19)   & 2.11      & $5.08\pm0.05$   & $11.40\pm0.10$            & -7.65\\
          & 0.42      & $4.69\pm0.11$   & $2.11\pm0.05$             & -3.03\\
(16-15)   & 1.92      & $6.03\pm0.09$   & $12.31\pm0.16$            & -7.76\\
(4-5)     & 2.56      & $5.56\pm0.06$   & $15.13\pm0.15$            & -7.81\\
(18-18)   & 2.10      & $5.14\pm0.08$   & $11.46\pm0.12$            & -7.82\\
          & 0.36      & $4.14\pm0.16$   & $1.57\pm0.06$             & -3.35\\ 
(12-13)   & 1.95      & $5.49\pm0.06$   & $11.41\pm0.15$            & -7.09\\
(23-23)   & 1.20      & $5.31\pm0.05$   & $6.79\pm0.05$             & -7.15\\
(17-17)   & 2.16      & $5.36\pm0.05$   & $12.30\pm0.11$            & -7.96\\
          & 0.18      & $2.62\pm0.22$   & $0.51\pm0.04$             & -4.44\\
(16-16)   & 2.13      & $5.30\pm0.05$   & $12.04\pm0.10$            & -8.22\\
          & 0.22      & $3.03\pm0.16$   & $0.74\pm0.04$             & -4.53\\
\hline
N6334~I(N) &           &                 &                          &            \\        
  (16-15) & 0.09      & 5.83$\pm$0.822  & 0.50$\pm$0.073            & -3      \\        
  (4-5)   & 0.42      & 4.46$\pm$0.188  & 1.88$\pm$0.680            & -3      \\        
  (18-18) & 0.13      & 6.07$\pm$0.841  & 0.73$\pm$0.096            & -3      \\ 
\hline 
I17233     &          &                 &                           &            \\        
  (19-19) & 0.49      & 8.73$\pm$0.639  & 4.58$\pm$0.296            & -5      \\        
  (16-15) & 0.47      & 9.47$\pm$0.440  & 5.10$\pm$0.206            & -5      \\        
  (4-5)   & 0.68      & 8.60$\pm$0.255  & 6.22$\pm$0.156            & -4      \\        
  (18-18) & 0.60      & 8.28$\pm$0.176  & 5.68$\pm$0.100            & -5      \\        
  (23-23) & 0.12      & 6.89$\pm$1.256  & 0.94$\pm$0.149            & -5      \\        
  (12-13) & 0.21      & 8.18$\pm$0.753  & 1.86$\pm$0.160            & -4      \\ 
\hline 
G9.62    &            &                  &                          &  \\
(21-21)  & 0.07       & $6.99\pm0.92$    & $0.52\pm0.06$            &  3.78\\
(20-20)  & 0.06       & $5.29\pm0.97$    & $0.36\pm0.05$            &  3.35\\
(19-19)  & 0.11       & $4.89\pm0.43$    & $0.56\pm0.04$            &  3.63\\
(16-15)  & 0.10       & $5.68\pm0.51$    & $0.64\pm0.05$            &  3.37\\
(4-5)    & 0.22       & $6.03\pm0.23$    & $1.40\pm0.05$            & 3.68\\
(18-18)  & 0.11       & $6.08\pm0.52$    & $0.71\pm0.06$            &  3.11\\
(12-13)  & 0.07       & $6.51\pm0.74$    & $0.48\pm0.05$            &  2.67\\
(17-17)  & 0.14       & $6.27\pm0.50$    & $0.90\pm0.06$            &  3.21\\
(16-16)  & 0.16       & $5.65\pm0.30$    & $0.98\pm0.05$            &  3.73 \\
\hline
G10.47    &           &                 &                           &            \\         %
(19-19) & 0.65      & 8.98 $\pm$0.727 & 6.21$\pm$0.442            & +66     \\         
(16-15) & 0.54      & 10.89$\pm$0.891 & 6.36$\pm$0.405            & +66     \\        
(4-5)   & 0.65      & 10.10$\pm$0.685 & 7.00$\pm$0.382            & +66     \\        
(18-18) & 0.68      & 9.90 $\pm$0.887 & 7.10$\pm$0.566            & +65     \\        
(23-23) & 0.29      & 10.08$\pm$1.651 & 3.13$\pm$0.378            & +66     \\        
(12-13) & 0.41      & 9.97 $\pm$0.911 & 4.41$\pm$0.345            & +66     \\  
\hline
G10.62    &           &                 &                           &          \\         
  (4-5)   & 0.29      & 7.65$\pm$0.495  & 2.28$\pm$0.126            & -3     \\  
\hline
G20.08N   &           &                 &                           &          \\         
  (4-5)   & 0.07      & 5.44$\pm$1.134  & 0.41$\pm$0.078            & +42    \\ \hline  
G29.96    &           &                 &                           &          \\        
  (19-19) & 0.19      & 6.10$\pm$0.506  & 1.17$\pm$0.081            & +98     \\        
  (16-15) & 0.16      & 6.61$\pm$0.543  & 1.13$\pm$0.080            & +98     \\        
  (4-5)   & 0.29      & 6.09$\pm$0.421  & 1.61$\pm$0.082            & +98      \\       
  (18-18) & 0.19      & 7.37$\pm$0.856  & 1.34$\pm$0.109            & +98     \\        
  (23-23) & 0.06      & 3.09$\pm$1.101  & 0.22$\pm$0.066            & +98     \\        
  (12-13) & 0.14      & 4.67$\pm$0.690  & 0.58$\pm$0.076            & +98     \\ \hline 
G31.97MM1 &           &                 &                           &           \\         
  (4-5)   & 0.12      & 4.46$\pm$0.616  & 0.56$\pm$0.065            & +97     \\  \hline     
G34.26    &           &                 &                           &           \\        
  (19-19) & 0.78      & 6.97$\pm$0.332  & 5.40$\pm$0.233            & +59      \\        
  (16-15) & 0.69      & 6.28$\pm$0.278  & 4.73$\pm$0.180            & +59      \\        
  (4-5)   & 0.86      & 6.44$\pm$0.222  & 6.22$\pm$0.182            & +59      \\        
  (18-18) & 0.82      & 7.67$\pm$0.623  & 6.22$\pm$0.462            & +59     \\         
  (23-23) & 0.28      & 7.07$\pm$0.841  & 2.30$\pm$0.220            & +58     \\         
  (12-13) & 0.49      & 5.55$\pm$0.661  & 2.00$\pm$0.194            & +59     \\  \hline 
W48    &           &                 &                           &           \\        
  (4-5)       & 0.04     & 4.54$\pm$0.993   & 0.19$\pm$0.038           & +42.34     \\        %
\enddata
\end{deluxetable}

\subsection{Null detection of maser emission?}

\begin{figure*} 
 \centering
 \includegraphics[angle=0, width=0.80\linewidth]{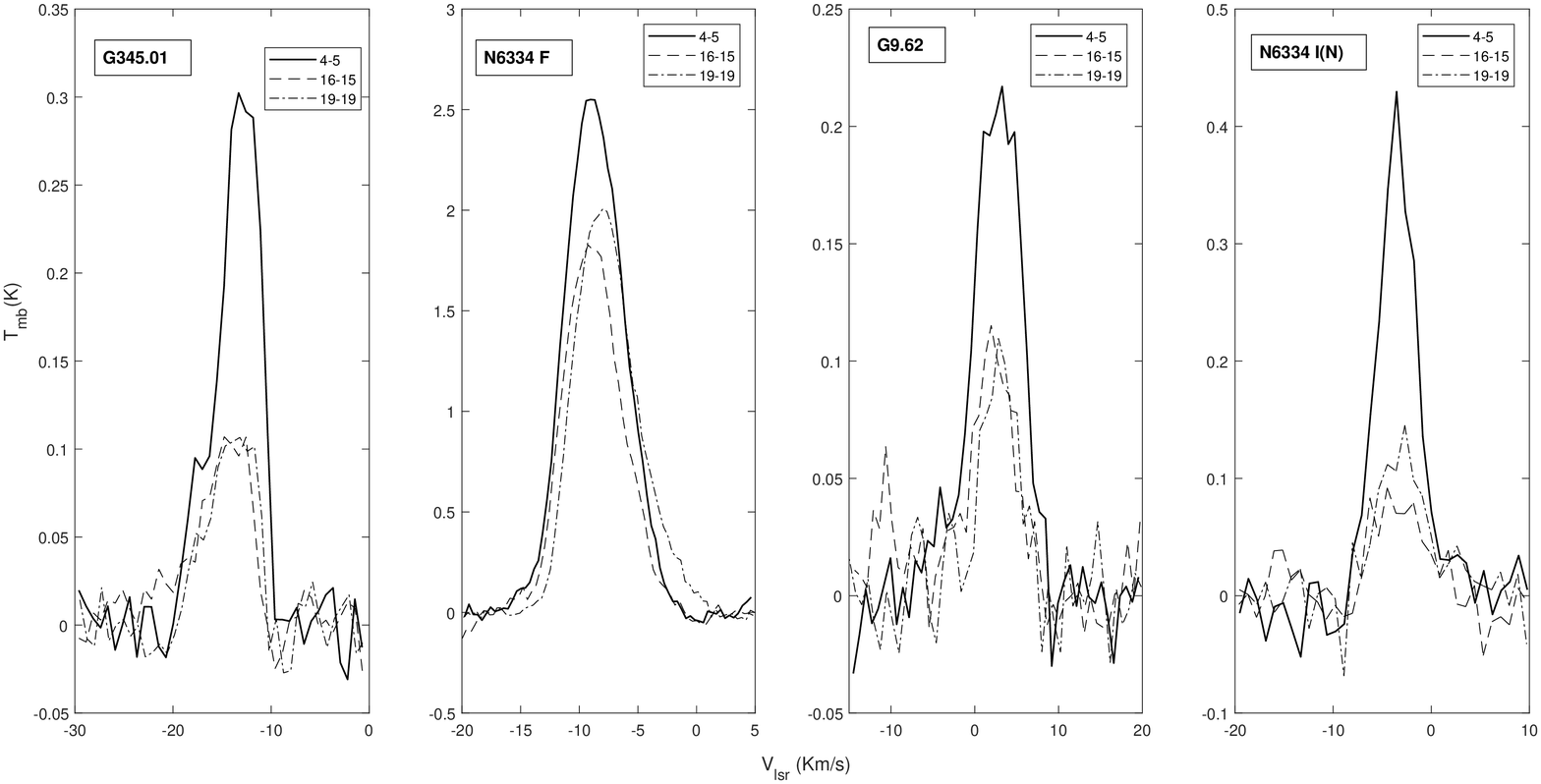} 
 \caption{
Spectra of methanol lines of G345.01, NGC6334F, G9.62, and NGC6334I(N). 
Thick solid lines shows the profiles of the $4_2-5_1$A$^+$ line, thin dashed lines, profiles
of two other methanol lines detected in these sources.
\label{allspec}}
\end{figure*}

The $4_2-5_1$~A$^+$ methanol line at 247.2\,GHz is a candidate to show
maser emission
\citep[][]{1997MNRAS.288L..39S,2005MNRAS.360..533C}. However, we
detected only broad lines at this frequency,  ranging from 4.5 to
  10.1~km~s$^{-1}$.  The similarity of the $4_2-5_1$~A$^+$ line profiles to the
  other thermal methanol lines (see Fig.~\ref{allspec})
 suggests that the former transition is also thermal. 
In addition, rotation diagrams constructed for eight
sources (see below) show that the ratios of different line intensities
correspond to local thermodynamic equilibrium (LTE). Thus, in most
cases, the emission in the 247.2\,GHz line is  probably thermal.

A possible exception is the spectral feature 
at $-17.62$~km~s$^{-1}$ in G345.01 (see Table~\ref{paramobs} and Fig.~\ref{allspec}).
The origin of this feature is uncertain. On the one hand, similar, but less
prominent features can be seen in the spectra of other lines (Fig.~\ref{allspec}).
Maser velocities in this object are usually more negative than 
$-20$~km~s$^{-1}$~\citep[e.g.][]{2012ApJ...759L...5E}. These facts 
indicate that the emission at  $-17.62$~km~s$^{-1}$ may be thermal.
On the other hand, \citet{2013MNRAS.433.3346K} detected a weak 
methanol maser in the 19.9~GHz $2_1-3_0$~E line in this source precisely 
at $-17.5$~km~s$^{-1}$, and~\citet{2018MNRAS.480.4851E} found  weak maser
emission in the $7_{-2}-8_{-1}$~E, $6_2-5_3 {\rm A}^-$, and $6_2-5_3 {\rm A}^+$ Class II 
transitions at 37.7, 38.3, and 38.5~GHz at around $-17$~km~s$^{-1}$ 
(in addition to stronger emission at velocities $< -20$~km~s$^{-1}$).
Interferometric observations with ALMA or the SMA could help establish
the origin of this feature.

G10.62, G31.97MM1, and NGC~6334~I(N) show fairly strong $4_2-5_1{\rm A}^+$ lines,
while other methanol lines are weak or absent. Because the 247.2\,GHz
lines are broad in these sources, most likely they are thermal and the
weakness/absence of other methanol lines means that the gas is
too cold to excite them. Note that
the upper energy level of the $4_2-5_1{\rm A}^+$ transition corresponds to 70 K,
while the upper levels of the other methanol lines in our passband
range from  338 to 661\,K. 

Because our source list contained the most promising sources from the viewpoint
of detecting masers in the $4_2-5_1$~A$^+$ transition (see~Section~\ref{sec:intro}),
we conclude that maser emission at 247.2~GHz, comparable in intensity
with that at 157~GHz or 107~GHz, either never arises, or only
arises in very rare cases; hence, the maser regime described by Model 3 
of \citet[][]{2005MNRAS.360..533C} is unusual for Class II masers.

Model 3 is characterized by a low gas kinetic temperature. While the dust,
whose radiation inverts the Class~II transitions, has a temperature of 175~K
in this model, similar to those in other models,
the gas temperature here is only 30~K. This is the lowest gas temperature model 
of \citet[][]{2005MNRAS.360..533C}. The model also predicts bright masers 
at 19.9 and 37.7~GHz, as well as weaker masers at  38.3, and 38.5~GHz --- which
are detected in G345.01. The second lowest gas temperature 
model (Model 7) also predicts a fairly high brightness temperature 
at 247.2~GHz, of $10^{5.6}$\,K. Note that the latter model does not predict bright masers
at 157~GHz. Other models have gas temperatures of 120~K and higher and do not predict
substantial emission at 247.2~GHz. Thus, masers at 247.2~GHz may exist only in cold
gas permeated by much hotter radiation. Our results show that such
conditions are unusal for star-forming regions, although they may occur in rare cases,
such as G345.01+1.79.

\subsection{Methanol detections in our sample}

CH$_3$OH~is a common molecule in MSFRs, with abundances of order
$10^{-9}$ in the cold gas phase \citep[e.g.][]{2000A&A...361..327V}.  Its formation is thought
to occur by grain-surface chemical reactions during an early, cold
stage in massive cores.  Gas-phase
production of CH$_3$OH~at temperatures $<100$\,K is mainly through
radiative association of CH$_3^+$ and H$_2$O and subsequent 
dissociative recombination of the resulting CH$_3$OH$_2^+$ ion. This pathway is
relatively inefficient, however, yielding abundances of only
$\sim10^{-11}$ relative to H$_2$~\citep{2006FaDi..133..177G}. Thus, gas-phase
chemistry cannot account for the observed abundances and grain surface
chemistry is invoked to reach the larger observed
abundances~\citep[e.g.,][]{2007A&A...467.1103G}.  Subsequent to its formation
on grain surfaces, methanol sublimates as the grains are heated during
the star formation process, either by radiation \citep[e.g.,][]{2000A&A...361..327V}  
or from shocks \citep[][]{Bachiller+1995,Bachiller+1998,2002ApJ...567..980G,2007ARep...51...44K}. 
Thus, the presence or absence of methanol emission, and its abundance, are expected 
to depend on the evolutionary state of the region.

 \begin{figure}
\centering
  \includegraphics[width=0.8\linewidth]{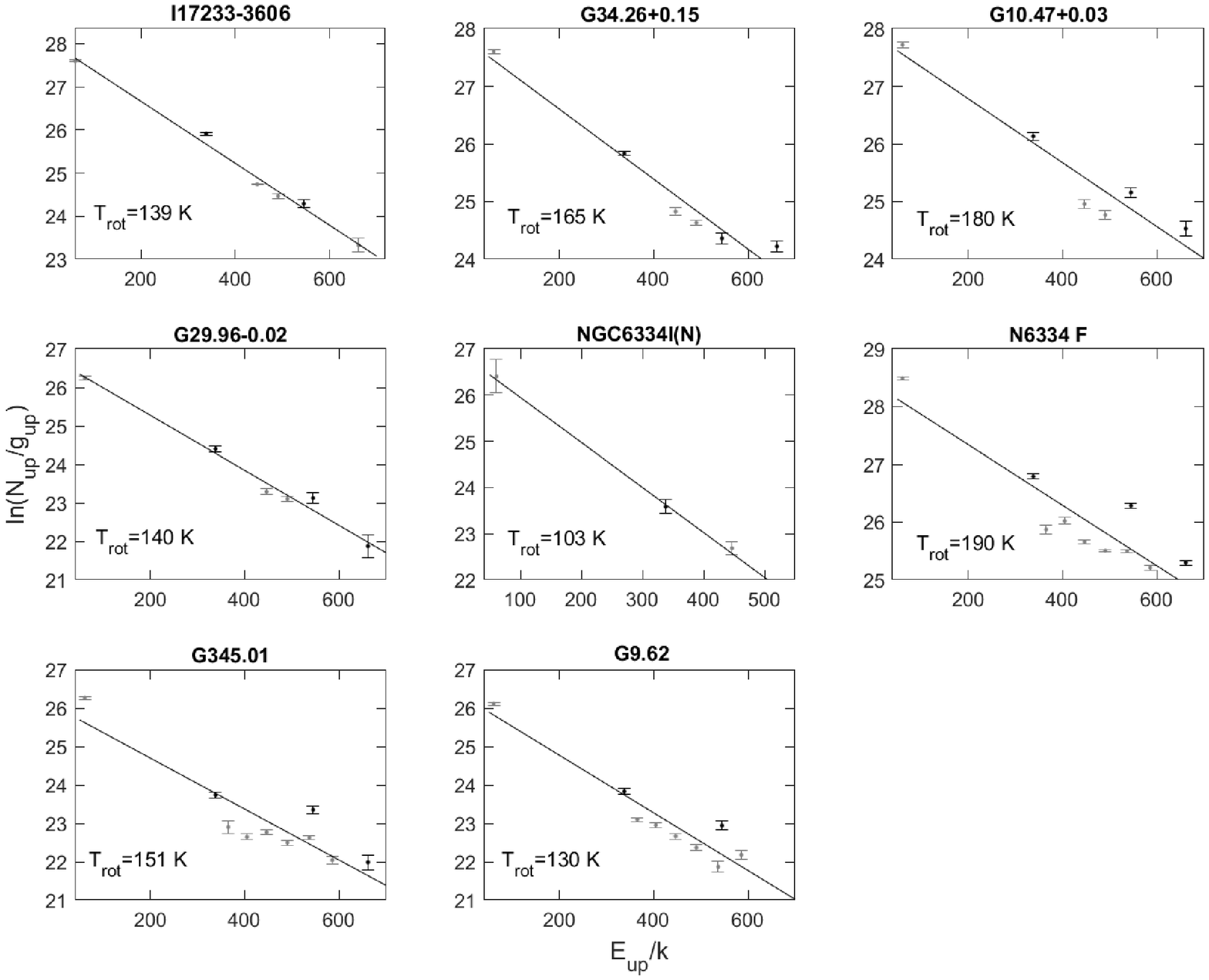}
  \caption{Rotation diagrams for methanol emission from eight sources.
The solid lines show the linear fits to all the
data points in the plots. The grey circles show A-methanol
levels and the black circles show E-methanol levels. 
} 
\label{rotdiag}
\end{figure}

 \begin{figure}
\centering
  \includegraphics[width=0.3\linewidth]{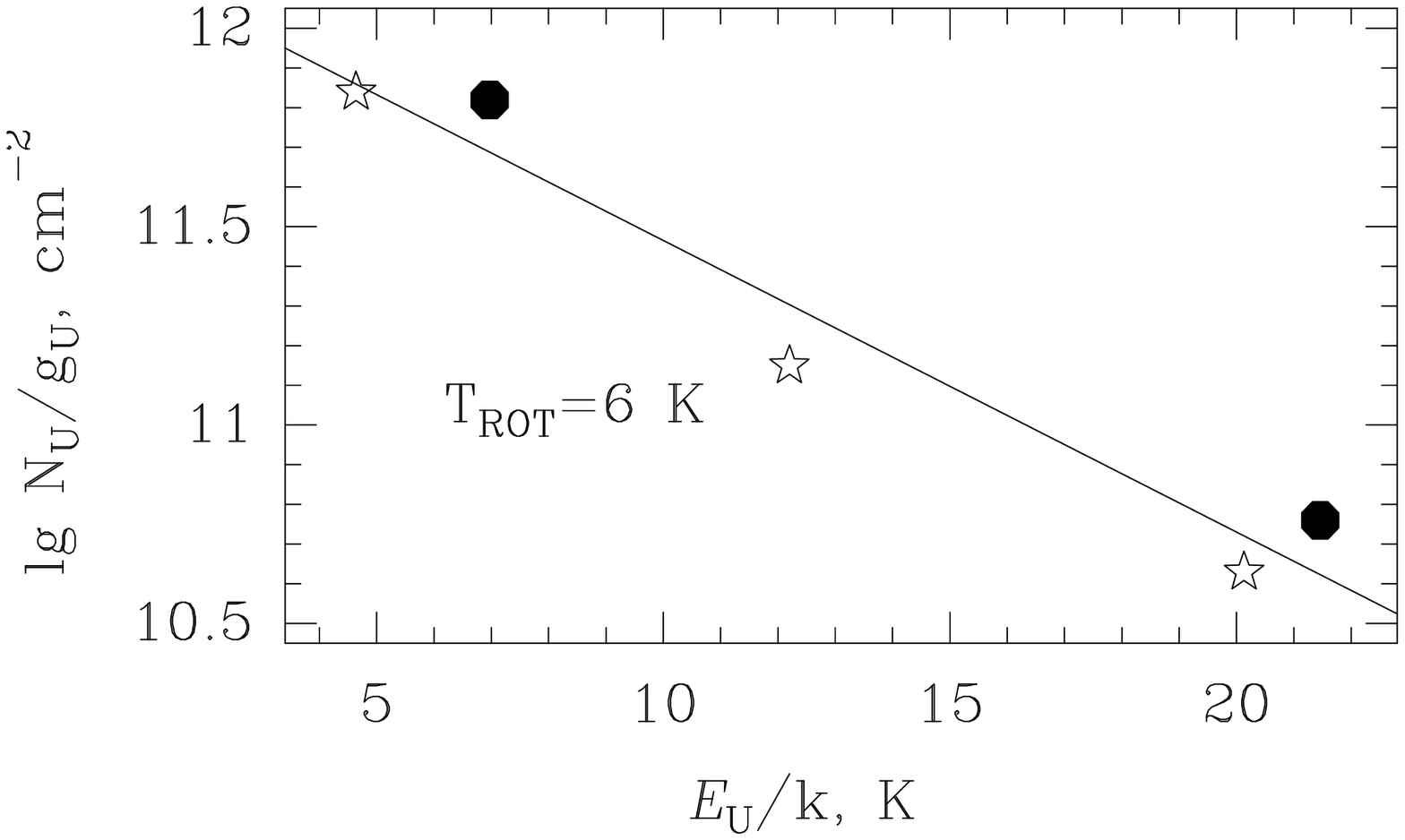}
  \caption{Model rotation diagram for A and E methanol. The filled circles show A-methanol
levels and the open stars show E-methanol levels.} 
\label{rotchen}
\end{figure}

The sixteen MSFRs we observed are in different evolutionary stages.  Table
1 indicates the general stage of each region, ranging from young (warm 
core), 
to advanced (hot
molecular core plus UC\,H{\small II}{} region).  We note that the HMC + UC\,H{\small II}{}
category also contains a range of evolutionary states:  in some cases
the UC\,H{\small II}{} regions are well-developed \citep[e.\,g.{} G34.26;][]{1989ApJS...69..831W} 
while in other cases they appear to be nascent \citep[e.\,g.{} NGC~6334~I(N);][] 
{2014ApJ...788..187H}.  Of the four warm cores, only
one (G31.97MM1), shows methanol emission. 

Among the eleven hot molecular
cores, there are notable differences in their methanol emission. 
Four of the HMCs (I17233, NGC6334F, G34.26, and G10.47)
are very rich in lines of methanol and other molecules, showing at least six
methanol transitions. Five other objects, G345.01, NGC~6334~I(N), G9.621, G10.62, 
and G29.96, have weaker lines of methanol and other molecules.
The remaining two HMCs (G45.07 and G20.08N) show
almost no methanol emission --- only a weak detection of the $4_2-5_1{\rm A}^+$ line
in G20.08N. 

\subsection{Rotation diagram analysis}

The CH$_3$OH~transitions we detect can be used to estimate the rotation temperature 
and column density of the gas via a rotation diagram (RD) analysis \citep[e.\,g.][]{1991ApJS...76..617T}. 
This method assumes that multiple transitions of a molecular species are optically 
thin and that the molecules are in LTE; that is, the ratios of all upper-level 
populations correspond to a single rotational temperature.
No background radiation is considered, and uniform temperature and density are 
assumed. If the molecular transitions are thermalized by collisions, the rotation temperature 
will closely approximate the gas kinematic temperature. 

Thus, the velocity-integrated main-beam brightness temperature, $\int T_{\rm mb}
dv$, is related to the species column density in the upper level, $N_{\rm u}$, by

\begin{equation}
{\rm ln} \left( \frac{3k\int T_{\rm mb}dv}{8\pi^3\mu^2\nu {\rm S}} \right) = {\rm ln} \left( \frac{N_u}{g_u} \right) = {\rm ln} \left( \frac{N_{\rm tot}}{Q_{\rm rot}}\right) - \frac{E_{\rm u}}{kT_{\rm rot}}
\label{eq:rdeq}
\end{equation}
where $\nu$ is frequency, $\mu$ the dipole moment, S is the line
strength, $g_{\rm u}$ is the total degeneracy of the upper state,
$N_{\rm tot}$ is the column density of the molecule in all states, and
$E_{\rm u}$ is the energy of the upper level. For the spectroscopic
parameters we used the values from the Cologne catalog~\citep{2001AA...370L..49M}. 
The partition functions of methanol, $Q_{\rm rot}$, that depend
on temperature, were derived from the catalog values.
The catalog provides partition functions for
a discrete sample of temperatures $T_{\rm cat}$; hence we had to interpolate between the
catalog values to obtain the numerical results appropriate for our temperatures. To calculate 
$Q_{\rm rot}$ for a temperature $T_{\rm rot}$, we used the catalog value of
the partition function for $T_{\rm cat}$ that is closest to $T_{\rm rot}$,
and multiplied it by $(T_{\rm rot}/T_{\rm cat})^{1.5}$.

Equation \ref{eq:rdeq} is a linear equation with slope ($-1/T_{\rm rot}$) and 
intercept ln($N_{\rm tot}/Q_{\rm rot}$). The rotation
diagram is the plot of ln($N_u/g_u$) versus $E_{\rm u}/k$ for each
molecular transition. The temperature and column density are then
determined via a linear least-squares fit. As we use beam-averaged
radiation temperatures the derived column densities are also
beam-averaged.

We label in Figures \ref{sp2010} and \ref{sp2017} the six to 
ten CH$_3$OH~transitions used in the linear fit.  Also 
labeled are the SO$_2$ and $^{34}$SO transitions. The resulting 
CH$_3$OH~rotation diagrams for the nine sources where
three or more methanol lines are detected are shown in Figure \ref{rotdiag}. 
The rotational temperatures obtained range from 104 to 190\,K, and column densities 
from $1.8 \times 10^{15}$ to $3.2 \times 10^{16}$ cm$^{-2}$; the
values are listed in Table 4. These column 
densities, of order $10^{15}$--$10^{16}$
cm$^{-2}$, assume a source size equal to the beam size of 25$''$.  More realistic
source sizes, taken from the literature, suggest column densities about two orders
of magnitude higher, ranging from $3.3 \times 10^{16}$ to 
$7.0 \times 10^{18}$ cm$^{-2}$. The fact that for each source
the positions of all points can be fitted by a single line corresponding
to a temperature higher than 100~K demonstrates that in all nine sources,
methanol emission is dominated by hot components in all observed lines,
including $4_2-5_1 {\rm A}^+$.


Note that a hot component of methanol emission in N6334F, N6334I(N), I17233, 
G10.47, and G34.26 have been analysed by~\citet{2017A&A...603A..33G} using high-energy 7--6 methanol 
lines at $\sim 338$~GHz. Our temperatures are roughly in agreement with those obtained 
by~\citet{2017A&A...603A..33G} for 
all sources except N6334I(N), where our temperature is lower by a factor of 1.7.
Note that the rotational temperature for this source was determined using only three
lines with the lowest energy levels in our sample, which may affect the result.
The temperature of 179.6~K, obtained by~\citet{2017A&A...603A..33G} using
a larger number of lines is probably more correct than our temperature.

To determine molecular column densities in hot cores from the results of single-dish
observations, hot core sizes are necessary. 
While~\citet{2017A&A...603A..33G} determined hot core source sizes indirectly, by fitting
the observed line intensities, we relied upon sizes taken from the literature; these were
measured directly, but using dust or molecules other than methanol. The sizes we use for
G34.26 and N6334I(N) are significantly larger (2.4 and 6.1 times,
respectively), and column densities are 4.5 and 49 times lower than the Giannetti et al. values. 
Both of these methods to obtain source sizes have obvious shortcomings, interferometric observations of high-energy
methanol lines are necessary to correctly determine methanol column densities.

Methanol is thought to form on grain surfaces via a sequence of
hydrogenation reactions: 

CO$\;\rightarrow\;$HCO$\;\rightarrow\;$H$_2$CO$\;\rightarrow\;$CH$_3$O$\;\rightarrow\;$CH$_3$OH. 

\noindent
\citet[][]{2004ApJ...616..638W} and \citet[][]{2004ApJ...614.1124H} found that the 
efficiency of this process depends on both the dust temperature and the composition 
of the grain surface.  Considering pure CO, as well as CO-H$_2$O 
and CO-H$_2$CO mixtures for the surface layer, they report that CH$_3$OH production 
can be relatively efficient for dust temperatures up to about 20~K. 
The hydrogenation efficiency dropped markedly above this temperature.

The initial A and E methanol abundance ratio, [A]/[E], also depends 
on the dust temperature \citep[e.\,g.{}][]{2011A&A...533A..24W}.
It is about 1.4 for a dust temperature of 10~K, tending toward unity with
increasing temperature; it is nearly unity at temperatures of 20~K and above.
There are no allowed radiative or collisional transitions between A
and E methanol, so these processes cannot alter the initial [A]/[E]
ratio. Proton exchange reactions with H$^+_3$ and HCO$^+$ can in
principle equalize the A and E methanol abundances, but these
reactions are quite slow \citep[e.\,g.{}][]{2011A&A...533A..24W}; hence the initial
[A]/[E] ratio should be preserved.

An over-abundance of A methanol was found in dark clouds by \citet{1988A&A...195..281F},
but the situation in massive star formation regions is still uncertain.
\citet{2011A&A...533A..24W} determined [A]/[E] ratios in seven MSFRs; 
in only two of these were the ratios consistent
with low formation temperatures (10 and 16~K). For a sample of
Extended Green Objects,  \citet{2013ApJS..206...22C} found a mean [A]/[E] ratio of
 1.66, which is consistent with a methanol formation temperature lower
 than 10~K.  However, the ratios for the individual sources from their sample 
range between 0.57 and 2.73. 

We note that \citet{2013ApJS..206...22C} assumed LTE in their analysis.
They built rotation diagrams using both A-type and E-type lines
of the $2_K-1_K$ series, and then, using the derived rotational temperature, 
calculated methanol column densities for each A-type and E-type transition.  
Finally, they averaged column densities for A and E transitions separately
and used the average column densities to calculate the [A]/[E] ratios.

However, methanol energy levels are rarely thermalized in the ISM 
\citep[e.\,g.][]{Kalenskii..Kurtz} and neglecting deviations from LTE 
can lead to erroneous [A]/[E] ratios. To demonstrate this, we built
a model rotation diagram applying LVG intensities of the same $2_K-1_K$ 
methanol lines as used by \citet{2013ApJS..206...22C}. 
The LVG calculations were performed with the RADEX
code \citep[][]{2007A&A...468..627V} for equal abundances of A and E methanol,
with the gas temperature set to 30~K and the density to $3\times 10^5$~cm$^{-3}$.
The resulting diagram is shown in Figure~\ref{rotchen}. The filled circles show
A-methanol lines, while the open stars represent E-methanol lines. Note that 
the points do not lie on a single straight line, thus demonstrating a notable 
deviation from LTE. The points corresponding to A-methanol transitions
are located above the approximating line, while the points corresponding 
to E-methanol transitions fall either on the line or below it. This means 
that if we assume LTE when estimating the column densities of A and E methanol, 
the former value will be higher than the latter one. 
The method used by~\citet{2013ApJS..206...22C} yields [A]/[E] equal to 2.1.


Thus, neglecting deviations from LTE can lead to erroneous
results. Unfortunately we could not analyze our results with
RADEX. This code uses the LAMDA molecular database which is limited to methanol
transitions between levels with $J\le 15$. The complete solution of how to correctly determine the
[A]/[E] ratio requires a special study and is beyond the scope of the
present paper.  \citet{Kalenskii..Shchurov} used a combination of the
rotation diagram method and statistical equilibrium calculations to
estimate the abundances of A and E methanol in the massive star
formation region L379~IRS1. The abundances in that case proved to be equal.

Inspection of Figure~\ref{rotdiag} shows that the points corresponding to both 
A and E methanol are well-fit by the same straight line for I17233,
G34.26, G29.96, and N6334I(N). The natural explanation of this is as follows: 
(a) methanol level populations are in LTE;  and (b) the abundances 
of the two species in our source sample are approximately equal. In the remaining
five sources, the points corresponding to methanol E lie slightly above the points corresponding to methanol A. We
believe that this is an excitation effect since there is no way to make E methanol 
more abundant than A methanol.  These results
assume that either the composition of grain mantles allows methanol formation 
at temperatures about 20~K or higher or that some process(es) can equalize 
gas-phase A and E methanol abundances faster than the proton exchange reactions studied
by \citet{2011A&A...533A..24W}.

\subsection{Discussion}

The methanol lines in our passband are listed in Table~2. Nine of these lines
have upper-level energies above 330 K and are detected only in
very hot regions. Therefore it is not surprising that 
the strongest emission in these lines was detected
in the most prominent hot cores of our sample. Rotation
temperatures of about 130~K and higher (see Table~\ref{physparam})
confirm that the methanol emission in these lines
arises from hot cores.  

Using source sizes from the literature and beam-averaged methanol column 
densities from our rotation diagram analysis (see Table~\ref{physparam}),
we obtain methanol column densities corrected for beam dilution. We use 
a beam dilution factor $\eta_{bd}=\theta_s^2 / (\theta_b^2 + \theta_s^2)$, 
where $\theta_b$ is the FWHM of the APEX beam and $\theta_s$ is the source size. In this way, 
methanol column densities vary from $3.3\times 10^{16}$~cm$^{-2}$ for NGC~6334~I(N)
to $7.0\times 10^{18}$~cm$^{-2}$ for G10.47. Column densities of H$_2$
for IRAS17233 and G10.47 were reported by~\citet{2014ApJ...786...38H};
using their values, we
 obtain methanol abundances of $0.5\times 10^{-7}$ for IRAS17233 and
$1.1\times 10^{-6}$ for G10.47. Liu et al. (2011,2013) reported hot core masses
for G9.62 and G34.26 (30 and 76\;M$_\odot$, respectively.) From these values we estimated hydrogen
column densities and methanol abundances of $1.8\times 10^{-7}$ for
G9.62 and  $3.7\times 10^{-7}$ for G34.26. Such methanol abundances are
typical for hot cores \citep[e.\;g.][]{2000A&A...361..327V}.

\begin{deluxetable}{lccccccc}
\tabletypesize{\scriptsize}
\tablewidth{0pt}
\tablecaption{Physical parameters of the observed sources \label{physparam}}
\tablehead{\colhead{} &\colhead{$T_{\rm rot}$}&\colhead{$N_{\rm CH_3OH}^c$} 
                                 &\colhead{$N_{\rm CH_3OH}^d$}
                                        &\colhead{$\Delta V^a$}&\colhead{$M_{\rm vir}^d$} 
                                            &\colhead{$\theta$}&\colhead{$N_{\rm CH_3OH}$/$N_{\rm H_2}$}\\
\colhead{Source} &\colhead{(K)} &\colhead{(cm$^{-2}$)}&\colhead{(cm$^{-2}$)}
        &\colhead{(km~s$^{-1}$)}&\colhead{(M$_\odot$)}&\colhead{(arcsec)}&\colhead{$\times 10^{-8}$}
}
\startdata
G345.01+1.79 & 151   & 2.1(15)   &           & 6.2      &     &        &\\
NGC6334F     & 190   & 3.2(16)   & $>$5.8(18)& 8.5     &$<215$& $<1.9^8$& $>9.7^e$\\  
NGC6334I(N)  & 104   & 1.8(15)   & 3.3(16)   & 5.0      & 290 & 6.0$^4$& 0.5$^e$\\   
I17233       & 139   & 1.3(16)   & 6.8(17)   & 8.4      & 149 & 3.5$^1$ & 5.0\\   
G9.62        & 130   & 1.8(15)   & 4.2(17)   & 5.9      & 145 & 1.6$^7$ & 18\\   
G10.47       & 180   & 1.9(16)   & 7.0(18)   & 10.0     & 554 & 1.3$^1$& 105\\  
G10.62       &       &           &           & 7.6$^b$  & 594 & 4.1$^1$ &\\
G20.08N      &       &           &           & 5.4$^b$  & 720 & 4.0$^5$ &\\
G29.96       & 140   & 3.2(15)   & 3.2(17)   & 5.7      & 212 & 2.5$^3$& 4.8$^e$\\   
G31.97MM1    &       &           & 1.5(15)?  & 4.4$^b$  & 841 & 17.0$^6$ &0.11\\  
G34.26       & 165   & 1.7(16)   & 6.8(17)   & 6.7      & 333 & 4.0$^2$& 37\\  
\enddata
\tablecomments{$^a$Averaged over all the detected
methanol lines. $^b$Using only the CH$_3$OH~$4_2{-}5_1A^+$ transition. $^c$Using
our beam size (25\rlap.{$''$}2) as the source size. $^d$Using source sizes
from the literature ($\theta$ column in this Table), with interferometric
observations of \citet{2014ApJ...786...38H}$^1$, \citet{2007ApJ...659..447M}$^2$.
\citet{2007AA...468..1045B}$^3$, \citet{2009ApJ...707....1B}$^4$, 
\citet{2013RAA....13.1295Y}$^5$, \citet{2017ApJ...849...25L}$^7$, 
\citet{2008AA...481..169B}$^8$, and $Spitzer$ data of Rathborne et al. 2006$^6$;
$^e$using the virial mass to calculate H$_2$ column density.}
\end{deluxetable}

Methanol abundances in the other hot cores were estimated  using the
deconvolved methanol column densities and H$_2$ column densities obtained
under the assumption that the hot core masses are equal to
their virial masses (see Section~\ref{vmass} below). The results are presented in 
Table~\ref{physparam}. The abundances are in the range $10^{-7}-10^{-8}$, i.e., 
typical hot core methanol abundances. Note that the measured masses are a factor
of a few lower than the virial masses (Sect.~\ref{vmass}), so the methanol 
abundances derived in this way are likely under-estimated by a factor of a few.

The four hot cores that exhibit the richest molecular 
spectra, NGC6334F, I17233, G10.47, and G34.26, all have methanol abundances about 
$10^{-7}$. Hot cores with poorer molecular spectra, NGC~6334~I(N),
G9.621, and G29.96, all have methanol abundances of a few$\times 10^{-8}$.

The remaining methanol line, $4_2-5_1$~A$^+$, has an upper-level energy of only 70 K 
and can arise in cooler, extended regions, surrounding the hot cores. Hence, this 
line was detectable in the less-prominent hot cores, G10.62 and G20.08. Three sources,
G345.01, NGC~6334~I(N), and G9.621, are an intermediate case. In these sources we 
detected emission in the 
$4_2-5_1$~A$^+$ line and also in the $16_2-15_3$~E and $18_3$~A$^- - 18_2$~A$^+$ lines, which 
have the lowest upper-level energies among the five ``hot core'' lines from our 
bandpass (338~K and 446~K, respectively). This is consistent with the fact that 
NGC~6334~I(N) had the lowest T$_{\rm rot}$ (104 K) of the eight sources for which 
we made rotation diagrams.

The four warm cores of our sample are embedded in IRDCs and have
masses ranging from about 400~M$_\odot$ to slightly over 2000~M$_\odot${}
\citep[][]{2006ApJ...641..389Rathborne}.  G24.33MM1 shows bright
8.0\,$\mu$m emission and has associated H$_2$O~and CH$_3$OH~maser
emission \citep[][]{2009ApJS..181..360C}.  G28.53MM2, G30.97MM1, and
G31.97MM1 are associated with green fuzzy cores (or extended
green objects), that show strong 4.5\,$\mu$m emission
\citep{2009ApJS..181..360C,2004ApJS..154..352N}.  G28.53MM2 has no
maser emission; G30.97MM1 has H$_2$O~and class~II (but not class~I) 
methanol masers;
while G24.33 and G31.97MM1 show H$_2$O~and both class~I and class~II
CH$_3$OH~ maser emission
\citep[][]{2009ApJS..181..360C,2012AN....333..634S}.  Of the four warm
cores, only G31.97MM1 shows weak 4$_2-5_1$~A$^+$ emission.

Estimation of methanol column density from the integrated intensity of
only one line and assuming LTE can lead to significant errors
\citep{Kalenskii..Kurtz}. Therefore, we estimated the
column density and abundance of methanol in G31.97MM1 in the following
way. First, we calculated the hydrogen volume density and column density, 
using the source size and mass from~\citep{2006ApJ...641..389Rathborne}.
These values were $1.3\times
10^6$~cm$^{-3}$ and $1.5\times 10^{24}$~cm$^{-2}$, respectively. Then
we computed several LVG models with the RADEX
code \citep[][]{2007A&A...468..627V}, using as input parameters the
calculated density and setting the kinetic temperature to 20~K 
(typical for IRDCs), and varying the methanol column density. Agreement
with the observed brightness temperature of the $4_2-5_1$~A$^+$ line was
achieved for a methanol column density of $1.5\times
10^{15}$~cm$^{-2}$. Calculating H$_2$ column density
from the cloud mass, presented in~\citet{2006ApJ...641..389Rathborne}, we obtained
a methanol abundance of $\sim 10^{-9}$, which is
typical for quiescent gas in dense cores of molecular clouds \citep[e.\,g.][]{2010ApJ...710..567Garay}.

From the above considerations, the four warm cores reported here
are probably in the early stages of star formation, but G31.97MM1 is
likely to be somewhat more evolved, showing 4.5\,$\mu$m emission, both
H$_2$O~and CH$_3$OH~maser emission, and weak 
CH$_3$OH~4$_2-5_1{\rm A}^+$ emission.  G28.53MM2, with no maser emission from any
species, is probably the youngest core.
G24.33MM1, with no line emission detected by us, but with 8.0\,$\mu$m,
H$_2$O~and CH$_3$OH~maser emission, is probably at an intermediate stage.

\subsection{Masses}

\subsubsection{Virial Masses}
\label{vmass}

The simplest form of the virial theorem,
\begin{equation}
\label{eq:vir1}
2T+W=0,
\end{equation}
describes a  cloud in equilibrium, neglecting contributions from magnetic fields and 
surface pressure. For a homogeneous spherical cloud, the internal kinetic
energy $T=3M\sigma^2/2$, the gravitational potential energy $W=-3GM^2/5r$, and
the one-dimensional velocity dispersion $\sigma = \Delta V/\sqrt{8\ln2}$.
Using Eq.~\ref{eq:vir1} one can calculate the virial mass of the cloud as
\begin{equation}
\label{eq:virm1}
M_{vir} = 5\sigma^2 r / G.
\end{equation}
Equation~\ref{eq:virm1}
can also be written in the more convenient form
\begin{equation}
\label{eq:virm}
M_{\rm vir} = 0.50~d~\theta~\Delta V^{2}
\end{equation}

\noindent
where $d$ is the distance, $\theta$ is the source
angular diameter, and $\Delta V$ is the line width, in\,kpc, arcsecond,
and km~s$^{-1}$, respectively. There are theoretical
and observational evidences that massive clumps are usually in
a state of near-virial equilibrium \citep[e.g.,][and references therein]{2018A&A...619L...7T}
and hence, virial masses can be used as the estimates of clump masses.

For $\Delta V$ we used the average line width
from the detected CH$_3$OH~lines, and for the source sizes we used the
literature values shown in the next-to-last column of Table~\ref{physparam}. 
In the case of G31.97MM1, we note
that the source size represents the extended core, while the sizes of
the other sources correspond to the hot cores. Hence, for G31.97MM1 we
are estimating the virial mass of the (larger) warm core; in all other
cases we are estimating the virial mass of the (smaller) hot molecular core.

The resulting virial masses are presented in Table~\ref{physparam}. For G31.97MM1
we obtained 841~M$_\odot$, which is about a factor of 2 less than the core mass of 
1890~M$_\odot$~reported by~\citet{2006ApJ...641..389Rathborne}. Accordingly, the cloud should
collapse. However, we neglected the magnetic energy of the cloud, which 
may be of the same order as the kinetic energy~\citep{1999ApJ...520..706C};
so the cloud may be close to equilibrium. The virial masses derived for hot cores 
were several hundred solar masses.  Comparison of our virial mass estimates 
for hot cores I17233, G10.62, and G10.47 with
the measured masses presented in \citet{2014ApJ...786...38H}
shows that the virial mass of G10.62 is about a factor of 5 higher 
than the measured mass, while the virial mass of I17233 is higher than the measured mass
by an order of magnitude.  Only for G10.47 are the virial mass and the measured
mass similar to one another. The virial mass derived for  G9.621
is a factor of 5 higher than the hot core mass, measured by \citet{2011ApJ...730..102L},
and the virial mass derived for G34.26 is four times higher than the hot core mass, 
measured by \citet{2013ApJ...776...29L}. Thus, the virial mass
estimates are typically higher than the measured gas masses.
A similar result was
obtained by~\citet{2014ApJ...786...38H}, who found that
the hot cores from their sample had virial masses greater than the
gas masses by factors ranging from 1.1 to 29.6.

\subsubsection{Virial versus Measured Mass}

There are several possible ways to understand the above result. First, the
cores may be expanding, although there is no observational or
theoretical support for this.  Second, it may be that the hot cores
undergo local gravitational
contraction~\citep{2011MNRAS.411...65B}. In this case the cloud
behavior is governed by the conservation of energy ($T\sim|W|$) rather
than the virial theorem (Eq.~\ref{eq:vir1}), and the observed
linewidths may correspond to half of the virial mass. This model may
be applicable to sources such as G10.62, where the virial mass from this work is
higher than the measured source mass of \citet{2014ApJ...786...38H} by 
only a factor of a few.
Finally, it is possible that the hot cores are confined by external
pressure. The pressure necessary to stabilize a spherical cloud with 
mass $m$, radius $r$, and velocity dispersion $\sigma$ can be found
from:

\begin{equation}
\label{eq:mvirpr}
P_s = \frac{1}{4\pi r^3}\left(\frac{3}{2}m\sigma^2-\frac{Gm^2}{R}\right)
\end{equation}
\citep{2011AIPC.1386....9K}. 
It is difficult to estimate the external pressure in 
the general case, but if a core is embedded in a much larger 
spherical massive core with a density inversely proportional to the squared distance 
from the center, then the equation of hydrostatic equilibrium yields the following
expression for the pressure $P_g$ at the hot core boundary:
\begin{equation}
P_g \approx \frac{\Sigma_{mc} G}{12 r^2}\biggl( \frac{\Sigma_{mc} \pi R^2}{2}
+ \Sigma_{hc} \pi R r\biggr)
\label{eq:pres}
\end{equation}
where $r$ is the hot core radius, $R$ is the massive core radius,
$\Sigma_{hc}$ is the mean surface density of the hot core
($\Sigma_{hc}=m/\pi r^2$) and $\Sigma_{mc}$ is that of the massive
core. Using hot core masses and sizes from~\citet{2014ApJ...786...38H}
and applying Eqs.~\ref{eq:mvirpr} and~\ref{eq:pres}, we
find that these cores can be stabilized by massive cores with column
densities of a few$\times 10^{23}$~cm$^{-2}$. Such column densities
are typical for massive cores in star-forming
regions~\citep{2011MNRAS.411...65B}.  Therefore, the hot cores, in
principle, could be stabilized by the pressure of the external clouds.
Careful mapping of the massive cores is necessary to test this
possibility.

\section{CONCLUSIONS}
\label{sec:conclus}

Sixteen massive star-forming regions were observed in spectral lines of
methanol and other molecules at 247 GHz, using the APEX telescope with
an angular resolution of 25$''$. Eleven of the regions are hot
molecular cores (most with UC\,H{\small II}~regions) while the other four
are warm cores with signs of ongoing star formation,
embedded in infrared dark clouds. Nine hot cores show rich molecular
line spectra, although the strength of different species and
transitions varies from source to source.  In contrast, the warm
cores do not show significant molecular line emission.

One of the observed methanol lines, $4_2-5_1$~A$^+$ at 247.228 GHz, is a
candidate for class II maser emission, similar in intensity to
previously known $J_0-J_{-1}$~E masers. Although we detected this line in
the majority of the sources, there was no evidence that the emission
is masing. The only exception is a weak spectral feature
in G345.01, observed at the same velocity as 19.9, 37.7, 38.3, and 38.5-GHz 
masers.

Multiple CH$_3$OH~transitions were detected toward
nine of the cores. Eight of these were detected in a sufficient
number of transitions to use the rotation diagram method to estimate
rotation temperatures and column densities. The temperatures lie in
the range of 104--168~K and column densities from $3.3\times 10^{16}$ to
$7.0\times 10^{18}$\,cm$^{-2}$. Using the average CH$_3$OH~line parameters, we
estimate virial masses in the range 145 -- 720~M$_\odot$~for the hot 
cores and  841~M$_\odot$~for the warm
core G31.97MM1. The hot core virial masses proved to
be significantly higher than the measured gas masses.  We suggest that
these hot cores may be confined by the external pressure of a
surrounding molecular core.

\section*{Acknowledgements}
We are grateful to the anonymous referee for valuable comments.
This work was partially supported by grants from  UNAM/DGAPA project 114514
and a CONACYT fellowship to VH-H. LAZ acknowledges additional
support from DGAPA/UNAM and CONACYT. SVK was partially
supported by travel grants from the DGAPA project 114514 and 
the Coordinaci\'on de Investigaci\'on Cient\'\i{}fica of the UNAM.
This research has made use of 
NASA's Astrophysics Data System Bibliographic Services.

\software{RADEX \citep{2007A&A...468..627V}, GILDAS/CLASS  http://www.iram.fr/IRAMFR/GILDAS}







\end{document}